\documentclass[conference]{IEEEtran}
\IEEEoverridecommandlockouts
\usepackage{cite}
\usepackage{amsmath,amssymb,amsfonts}
\usepackage{algorithmic}
\usepackage{graphicx}
\usepackage{textcomp}
\usepackage{xcolor}
\usepackage{natbib}
\usepackage{booktabs}
\usepackage{multirow}
\usepackage{adjustbox}
\usepackage{subcaption}
\usepackage{graphicx}
\usepackage{url}
\def\BibTeX{{\rm B\kern-.05em{\sc i\kern-.025em b}\kern-.08em
    T\kern-.1667em\lower.7ex\hbox{E}\kern-.125emX}}
\begin{document}

\title{Parametric Retrieval-Augmented Generation using Latent Routing of LoRA Adapters\\
}

\author{
    \IEEEauthorblockN{Zhan Su$^1$, Fengran Mo$^2$\IEEEauthorrefmark{2}\thanks{\IEEEauthorrefmark{2} Corresponding authors.}, Jinghan Zhang$^3$, Yuchen Hui$^4$, Jia Ao Sun$^4$, Jian-Yun Nie$^4$\IEEEauthorrefmark{2}}
    \IEEEauthorblockA{$^1$School of Information Technology, Halmstad University, Halmstad, Sweden, zhan.su@hh.se}
    \IEEEauthorblockA{$^2$Department of Computer Science, Rochester Institute of Technology, Rochester, USA, frmvcs@rit.edu}
    \IEEEauthorblockA{$^3$School of Computing, Clemson University, Clemson, USA, jinghaz@clemson.edu}
    \IEEEauthorblockA{$^4$Department of Computer Science and Operations Research, Université de Montréal, Montreal, Canada, nie@iro.umontreal.ca}
}

\maketitle

\begin{abstract}

Parametric Retrieval-Augmented Generation (PRAG) is a RAG approach that integrates external knowledge directly into model parameters 
using a LoRA adapter, aiming at reducing the inference cost compared to traditional RAG. However, current PRAG approaches adopt a \textit{one-to-one} document encoding scheme, using a dedicated LoRA adapter for each individual document. This scheme introduces two major limitations:
1) As the number of documents increases, there will be a prohibitive cost for training and storage. 2) The LoRA adapters may largely overlap due to the shared knowledge across documents, making the approach highly inefficient.
To overcome these challenges, we propose the Poly-PRAG approach, which uses a small set of LoRA adapters that are able to encode more general knowledge. Each document can be encoded using a combination of them through a latent routing function.  
By jointly training the LoRA adapters and the \textit{latent routing function}, each LoRA adapter is able to encode a shared part of the knowledge across documents, and the routing function can select the best combination of adapters for a document.
Experimental results on four benchmarks demonstrate the effectiveness of the Poly-PRAG compared to other strong PRAG baselines. In addition, this approach reduces the storage requirement by avoiding the need to store a large number of LoRA adapters and offers a more efficient way to encode external knowledge into LLMs. The code is available\footnote{\url{https://github.com/shuishen112/poly-prag-encoding}}. 
\end{abstract}


\section{Introduction}
\begin{figure}
\small
    \centering
    \includegraphics[width=0.9\linewidth]{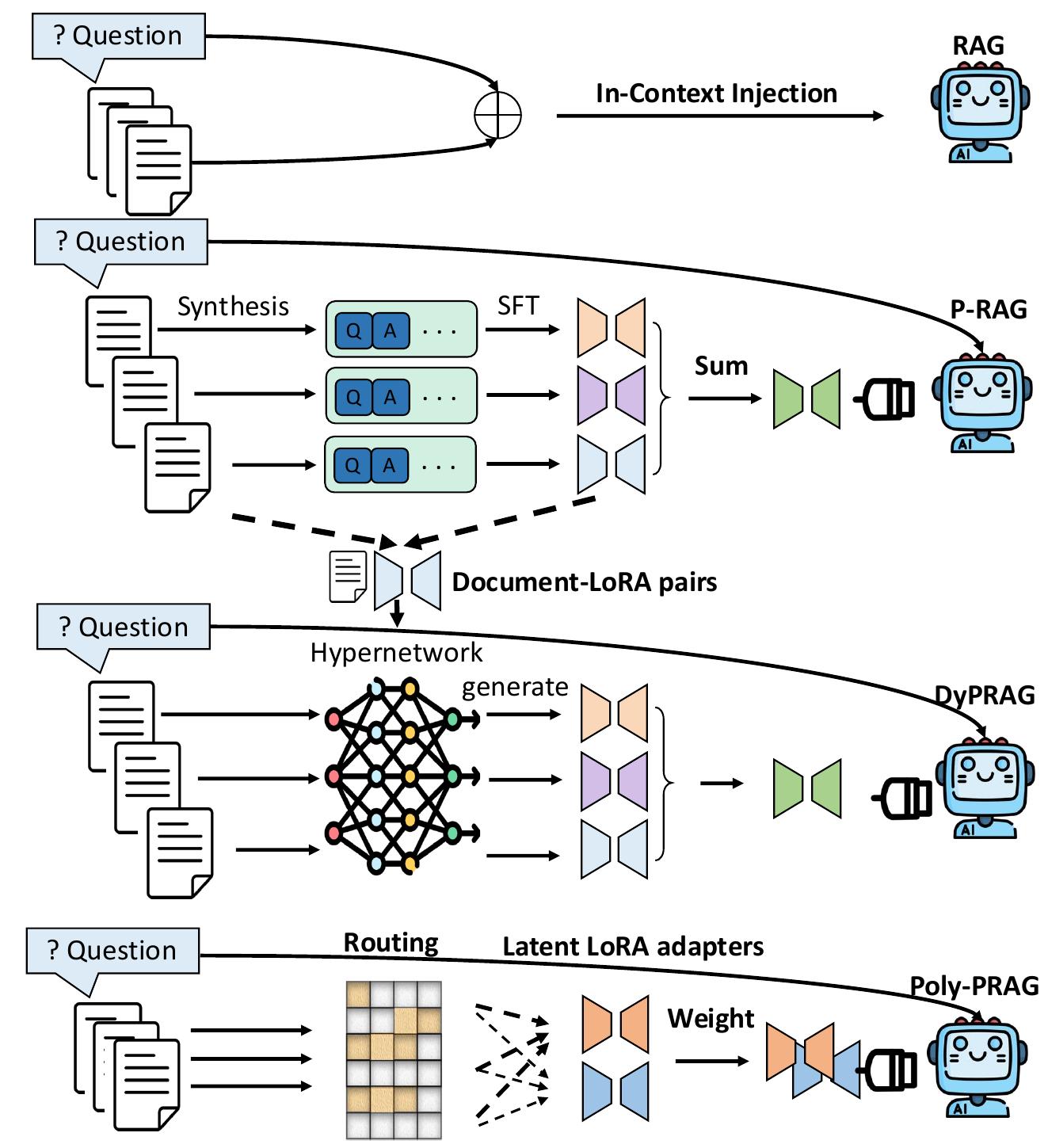}
    \caption{(1) Traditional RAG. (2) P-RAG creates one adapter per document. The sum of the adapters of the retrieved documents is added to LLM. 
(3) DyPRAG uses a hyper-network to translate individual documents to their corresponding LoRAs, and the average is injected into LLM. 
(4) Poly-PRAG trains a fixed number of adapters and a routing function to activate adapters for each document. The activated adapters are merged with weights and injected into the LLM.}
    \label{fig:methods_comparison}
\vspace{-3mm}

\end{figure}
Large Language Models (LLMs) have demonstrated impressive capabilities in various downstream tasks \cite{achiam2023gpt}. However, an LLM trained on 
a pre-training dataset may have difficulty answering questions requiring more up-to-date knowledge 
\cite{kandpal2023large,kaplan2020scaling}. To address this issue, Retrieval-Augmented Generation (RAG)~\cite{lewis2020retrieval} has emerged as a popular approach to allow LLMs to incorporate external knowledge from retrieved documents 
in order to enhance the answer generation process~\cite{guu2020retrieval,kandpal2023large,asai2024self,li2025lara,mo2026opendecoder}. 

The basic RAG method, \textbf{in-context injection}, retrieves a set of documents and injects them 
into the input prompt, letting the LLM incorporate them in answer generation  
\cite{shi2024enhancing,borgeaud2022improving,izacard2020leveraging,lewis2020retrieval,su2024dragin,zhang2025entropy}.   
To enhance the effectiveness of  RAG, existing studies have focused on optimizing the utilization of the retrieved documents during the generation phase
\cite{shi2024enhancing,jin2024tug,wang2024astute,weller2022defending,cho2024typos,yoran2023making,zhang2025ratt}. 
Despite some success, injecting knowledge via input prompts inevitably increases the context length, which not only introduces significant computational overhead but also hinders the performance of LLMs by introducing more noise into the context. 
The performance degradation has been observed due to the difficulty for LLMs to discern useful information within the increasingly noisy passages retrieved~\cite{shi2023large,yoran2023making,tu2025rbft}. 

To address this problem, a new approach, Parametric Retrieval Augment Generation (PRAG),
has been proposed~\cite{prag,tan2025dynamic} 
to integrate external knowledge through \textbf{parameters injection}. Specifically, an implementation of PRAG, noted as P-RAG \cite{prag}, leverages Low-Rank Adaptation (LoRA) adapters to parameterize each document \citep{hu2022lora}, which is then directly injected into the feed-forward layers of the LLM for generation. It is expected that the LoRA adapter can extract the important content from the document, and that injecting the resulting LoRA representation into the LLM incurs almost no additional inference cost because it can be merged with the original weights.
As shown in  Fig. \ref{fig:methods_comparison}, there are two stages for P-RAG: \textbf{offline encoding} and \textbf{online inference}. In the offline stage, P-RAG rewrites document $d_i$ into several question-answer pairs and trains the corresponding LoRA adapter for each document. 
In the online stage, given the query $q$, once a document $d_i$ is retrieved, its corresponding LoRA adapter will be merged into the LLM. 
  
While this \textbf{one-adapter-per-document} P-RAG approach succeeds in reducing inference cost compared to in-context injection, 
it suffers from several limitations. First, the large number of LoRA adapters required for a large document collection leads to prohibitively high training time and offline storage.  
Second, by treating each document as an independent island of information, the LoRA model may overfit to a single document while neglecting the common knowledge shared across documents, e.g., the shared terminology, recurring reasoning patterns, and conceptual bridges.  These limitations  raise the following questions: 
  \textit{Is it necessary to create a different adapter for each document? Can we create a set of LoRA adapters that capture some ``common knowledge'' 
  shared across documents?}
  If this is possible, then we only have to store a limited set of adapters for a large set of documents.
  
To answer these questions, we take inspiration from topic modeling 
\cite{blei2003latent}, which aims to address similar problems in language modeling: instead of building a probabilistic distribution of words for each document, a set of topics, each corresponding to a probabilistic distribution, is created, and a document is modeled as a mixture of topics.
A great advantage is that only a limited number of topic models are built and used for all documents through different mixtures. The LoRA adapters we want to build are intended to play a similar role to topic models, which are expected to capture shared knowledge within latent clusters of documents.

More concretely, we propose a novel document encoding method Poly-PRAG (Fig. \ref{fig:methods_comparison}), 
where we assume that there is a small set of base LoRA adapters that can encode a large set of documents (\textbf{many-documents-to-fewer-adapters}). To achieve this mapping, each document representation is associated with a mixture of latent LoRA adapters through a routing function \citep{ponti2023combining,page2023multi}. 
By jointly learning the LoRA adapters and the routing function, we expect that the adapters are able to capture the most important knowledge shared by some documents, while the routing function is able to assemble the adapters to encode the specific knowledge in a document.
This design is expected to be more 
  efficient in storage than the existing PRAG approaches because we only have to store a small set of LoRA adapters and the routing function.

One additional advantage may emerge due to the sharing of LoRA adapters across documents. In the existing PRAG approaches, a LoRA adapter is trained with the augmented data from a single document. In Poly-PRAG, a LoRA adapter may leverage the augmented data from several documents that share the adapter. We can expect better-trained LoRA adapters, therefore, increased effectiveness in the downstream tasks.
  
These expected advantages are confirmed in our experiments. 
We evaluate our method on QA tasks on the following datasets:  2wikimultihopQA (2WQA) \citep{ho2020constructing}, HotpotQA \cite{yang2018hotpotqa}, PopQA \cite{mallen2022not}, and ComplexWebQuestions\cite{talmor2018web}. The experimental
  results demonstrate that Poly-PRAG outperforms the existing strong PRAG baselines both in storage and inference efficiency and in effectiveness.

  The main contributions of this paper are as follows:

  \begin{itemize}
      \item We propose a novel parametric retrieval-augment generation framework (Poly-PRAG), which creates a compact set of LoRA adapters for the entire document collection, and each document is encoded by a combination of the adapters through a routing function.

      \item We perform a joint learning of LoRA adapters and the routing function so that specific knowledge of a document can be best captured.
      \item Experimental results on four benchmarks demonstrate both efficiency and effectiveness of Poly-PRAG compared to other PRAG baselines. 
      
  \end{itemize}

\section{Related Works}

Traditional retrieval-augmented generations (RAG) enhance the performance
  of LLMs by incorporating external knowledge by following a ``retrieve-then-read'' framework~\cite{borgeaud2022improving,guu2020retrieval,chen2024benchmarking}, which has demonstrated impressive capabilities in knowledge-intensive tasks. Building upon the RAG framework, a rich body of research work has been proposed to further improve the effectiveness \cite{shi2024enhancing,jin2024tug,wang2024astute,cho2024typos,jiang2024longrag}. For example, some studies focused on enhancing retrieval quality through query reformulation \cite{ma2023query} or document re-ranking \cite{sun2023chatgpt,yu2024rankrag}. 
 Another line of work focuses on improving how LLMs understand and use external documents \cite{jin2024tug, wang2024astute, weller2022defending, cho2024typos, yoran2023making}. Self-RAG \cite{asai2024self} introduces a self-reflection framework to improve generation quality and factual accuracy. It asks the model to judge whether the retrieved external information is useful for generation.  CRAG \cite{yan2024corrective} proposes a lightweight retrieval evaluator that assesses the quality of retrieved documents and injects external passages based on a confidence score. In these methods, external knowledge is usually incorporated into the model through input prompts. A critical problem is that the input may quickly become long, 
  leading to high inference costs. 
  
  To address this problem, parametric retrieval-augmented generation (PRAG) approaches are proposed \cite{prag, tan2025dynamic, tang2025role,chen2025privacy,su2026decoupling}, which encode a document using a LoRA adapter. The retrieved external knowledge is injected directly into the LLM’s feed-forward layers for generation.  
  The first PRAG approach -- P-RAG \cite{prag} builds a LoRA adapter for each document through augmented data (i.e., generated question-answer pairs). A critical problem is that a large number of adapters should be trained and stored, increasing the training and storage costs.
  Building on this, Dynamic PRAG (DyPRAG) \cite{tan2025dynamic} first encodes each document with one LoRA adapter as in PRAG. Then, a hypernetwork is trained to reproduce the LoRA adapter parameters, and the LoRA adapters can be removed after training. During online inference, DyPRAG uses the translator to generate the LoRA representation for a document.  DistilledPRAG \cite{chen2025privacy} extends DyPRAG and develops a new hypernetwork based on knowledge distillation. However, DyPRAG still needs to create a separate adapter for each document, incurring the same amount of training cost as the original PRAG approach.
  
  Unlike the one-adapter-per-document approach, our method assumes that a small set of LoRA adapters can be built, each encoding a distinct aspect of the documents in the collection. A document can be encoded by a mixture of the adapters selected by a routing function. By jointly training the latent routing function with the adapters, we expect that the specific content of each document can be adequately encoded. This approach can reduce the storage requirement for parametric document representations tremendously and improve the overall effectiveness of PRAG.

\section{Preliminaries}

\subsection{LoRA adapters}
LoRA is a parameter-efficient adapter architecture that achieves a competitive balance between performance and parameter efficiency \cite{hu2022lora}. For each linear transformation corresponding to the query ($q$), key ($k$), value ($v$), and output ($o$) of the self-attention layers, LoRA modifies the base model parameters as follows:
\begin{equation}
h = W_0 x + s\cdot A (B)^\top x
\label{eqn:lora}
\end{equation}
where $W_0$ are the (frozen) weights of the pre-trained base model. $A, B \in \mathbb{R}^{d \times r}$ are low-rank learnable parameters and $s\ge 1$ is a tunable scalar hyperparameter. 
\subsection{Parametric Retrieval-Augmented Generation}
\label{sec:pre_prag}

\textbf{Parametric RAG (PRAG)} represents an alternative approach to traditional RAG by integrating external knowledge directly into the parameters of the language model ($\mathcal{L}$) \cite{prag,tan2025dynamic}, aiming to circumvent the computational and memory overhead for long input contexts.
In this approach, each external document $\boldsymbol{d}_i \in K$ (where $K$ is the external knowledge corpus) is transformed into \textbf{parametric weights} $\pi_i = \mathcal{G}(\boldsymbol{d}_i)$, where $\mathcal{G}$ is a dedicated mapping function. To ensure an effective transformation that captures the document's content, PRAG employs a technique inspired by \textbf{document Augmentation} to generate diverse linguistic variations and knowledge-rich Q-A pairs. Specifically, the original document $\boldsymbol{d}_i$ is first rewritten into multiple variations $\{\boldsymbol{d}_i^1, \boldsymbol{d}_i^2, \ldots, \boldsymbol{d}_i^n\}$, and an LLM model 
is prompted to generate a set of question-answer (QA) pairs, resulting in the augmented dataset $\mathbf{A}_i$ for document $\boldsymbol{d}_i$:
\begin{equation}
    \mathbf{A}_i = \left\{\left(\boldsymbol{d}_i^k, \hat{q}_i^j, \hat{a}_i^j\right) \mid k \in [1, n], j \in [1, m]\right\}
\end{equation}
where 
$n$ is the number of document variations, and $m$ is the number of QA pairs per variation. Each generated triple $(\boldsymbol{d}_i^k, \hat{q}_i^j, \hat{a}_i^j)$ is concatenated to form a training sample $\boldsymbol{I} = [\boldsymbol{d}_i^k ;\hat{q}_i^j; \hat{a}_i^j]$.

As mentioned earlier, creating a LoRA adapter per document will incur a large storage cost. DyPRAG \cite{tan2025dynamic} is built on PRAG. A set of document-LoRA pairs $\mathcal{K}=\{(d_i,\text{LoRA}_i)\}$ is trained as in PRAG. It then uses a hypernetwork to translate a document to the parameters of the LoRA adapter. 
For a document $d_i$, DyPRAG extracts the last hidden state $s_i$ at the final token position as the input to the hypernetwork $\mathcal{F_{\phi}}$ to reproduce the parametric $\mathrm{LoRA}_i$ representation, i.e., $A_i$ and $B_i$ matrices.  

Despite the fact that the trained hypernetwork requires much less storage than the initial LoRA adapters, there is still a high cost for generating and storing them during the training process. In addition, generating a LoRA adapter from a partial input $s_i$ may reduce the capability of the translator to generate good LoRA parameters, thus leading to lower effectiveness than P-RAG (as confirmed in our experiments). 

\section{Poly-PRAG: Latent Routing for Document Parametrization}

\begin{figure*}
\small
    \centering
    \includegraphics[width=0.9\linewidth]{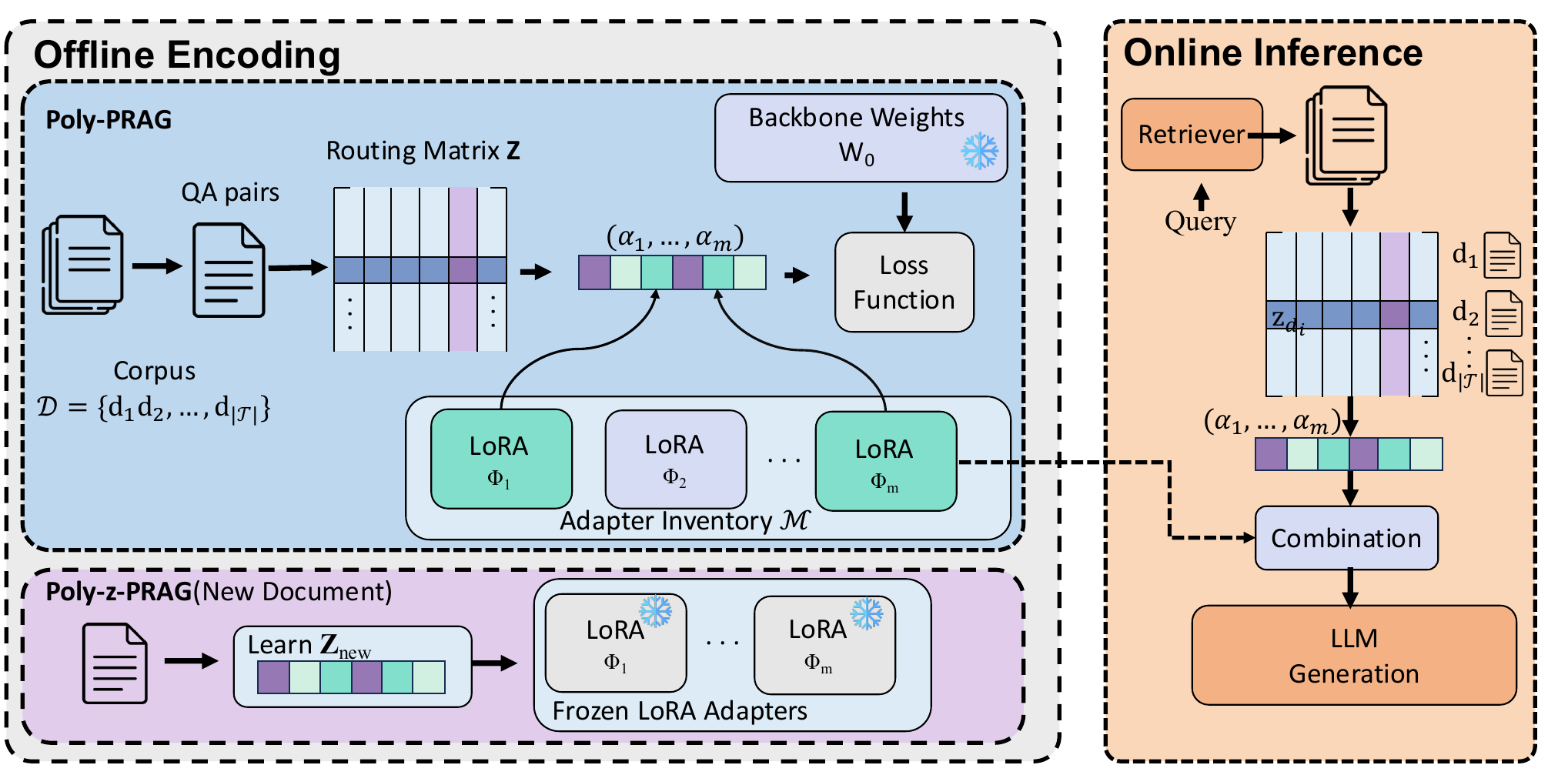}
    \caption{For the offline encoding (left), Poly-PRAG encodes the whole set of documents with a small set of latent LoRA adapters, and a routing function is used to select the adapters for each document. During inference time (right), for a given query and top-k retrieved documents, the routing function will select latent LoRA adapters to combine with the LLM for generation.}
    \label{fig:poly_prag}
\vspace{-4mm}
\end{figure*}

In this section, we present the Poly-PRAG framework, which is illustrated in Fig.~\ref{fig:poly_prag}. We first describe the shift from the one-adapter-per-document (one-to-one) encoding paradigm to a many-to-few paradigm (Section~\ref{sec:encoding_method}). We then explain how the LoRA adapters and a latent routing function are jointly trained during the offline encoding process (Section~\ref{sec:adapter_routing}). Finally, we introduce the online inference procedure of Poly-PRAG (Section~\ref{sec:online_inference}). 

\subsection{From One-to-One to Many-to-Few Paradigm}
\label{sec:encoding_method}


Inspired by the principle of topic modeling \cite{blei2003latent}, which creates a set of general topic models and uses a mixture of these models to model each document, we intend to create a set of general LoRA adapters playing the role of general encoders of some aspects of document contents. Each document can then be encoded by combining the general adapters in a specific manner.
Similar ideas have been used for generating task-dependent representations in MHR \cite{mhr}, or selecting relevant adapters for tasks in Poly \cite{ponti2023combining}. Our approach generalizes the idea and applies it to PRAG.

We posit that there exists a limited fixed inventory of base LoRA adapters $\mathcal{M}=\{\phi_1, \ldots, \phi_m\}$, where $|\mathcal{M}|\ll |\mathcal{T}|$. Each LoRA adapter is an expert for some encoding task, corresponding to an independent facet of knowledge parametrization across documents. Poly-PRAG aims to encode the whole document collection with a limited number of adapters: 
\begin{equation*}
    \{\boldsymbol{d}_1,\boldsymbol{d}_2,...,\boldsymbol{d}_{|\mathcal{T}|}\} \stackrel{\text{Poly-PRAG}}{\longrightarrow} \{\text{LoRA}_{\mathbf{1}},\text{LoRA}_{\mathbf{2}}, \text{LoRA}_{m}\}
\end{equation*}

We call this encoding schema many-to-few. 
Compared to the one-to-one encoding, there are several potential advantages for this many-to-few encoding: \textbf{1)} This method effectively compresses the knowledge of $|\mathcal{T}|$ documents into $|\mathcal{M}|$ shared adapters, drastically reducing the required storage. 
\textbf{2)} As the LoRA adapters are used to encode the whole document collection, they need to cover different aspects of document contents, so that collectively, they can encode the specific knowledge of each document. Therefore, the adapters are mutually complementary. 
\textbf{3)} As a LoRA adapter is shared by multiple documents, its training can benefit from the training data of multiple documents. There may be an increased benefit of data augmentation.

\subsection{Document-Aware Routing for Offline Training}
\label{sec:adapter_routing}

Given a set of LoRA adapters, each document is encoded by combining them through a routing function. This function depends on the LoRA adapters, and vice versa. Therefore, we design a joint training process
to train the routing function and LoRA adapters in the offline encoding. Specifically,  Poly-PRAG contains two components in each transformer layer: (1) an inventory of LoRA adapters $\mathcal{M}=\{\phi_1, \ldots, \phi_m\}$, where $\phi_i$ corresponds to $A^{(i)}, B^{(i)} \in \mathbb{R}^{d \times r}$, and (2) a routing function $r(\cdot)$ that combines these adapters for each document. Let $\mathcal{D}=\{\boldsymbol{d}_1,\boldsymbol{d}_2,...,\boldsymbol{d}_{|\mathcal{T}|}\}$ denote a corpus of  documents. We define a document routing function as the following matrix:

\begin{equation}
    r(\cdot)=\mathbf{Z} \in \mathbb{R}^{|\mathcal{T}| \times |\mathcal{M}|}
\end{equation}
Each row $\mathbf{z}_i = \mathbf{Z}[i,:]\in\mathbb{R}^{|\mathcal{M}|}$ corresponds to the routing logits (or probabilities) associated with document $d_i$. 

Unlike mixture-of-experts \cite{fedus2022switch}, which performs token-level top-$k$ routing, $\mathbf{Z}$ converges to a document-level routing, defining a soft partition over latent LoRA adapters. This is achieved by using a Gumbel-sigmoid distribution, which provides a continuous, differentiable relaxation of Bernoulli sampling, enabling gradient-based optimization for discrete binary variables during training \cite{jang2016categorical,ponti2023combining}, with  $\hat{\mathbf{Z}}_{i,j} \sim \texttt{Gumbel}(\mathbf{Z}_{i,j})$. Then the activated LoRA adapters are merged by the routing weights:
\begin{equation}
A^{\tau} = \sum_{i=1}^{m} \alpha_{i} A^{(i)}; \quad
B^{\tau} = \sum_{i=1}^{m} \alpha_{i} B^{(i)}
\label{eqn:poly-prag}
\end{equation}

The mixing coefficients are defined as $\alpha_i = \frac{\hat{Z}{\tau,i}}{\sum_j \hat{Z}{\tau,j}}$, where the active logits are normalized. This design allows to activate different LoRA adapters in each layer and combining them in a document-specific manner. Finally, the output of the Poly-PRAG layer is added to the output of the corresponding layer in the frozen backbone model ($W_0$):
\begin{equation}
    h = W_0 x + s\cdot A^\tau (B^{\tau})^{\top} x
\end{equation}

As a result, the overall objective is to optimize the loss function:

\begin{equation}
    \min_{\Delta \Phi, \Delta \mathbf{Z}} \sum_{(\boldsymbol{d}_i^k, \hat{q}_i^j, \hat{a}_i^j) \in \mathbf{A}_i} \sum_{t=1}^T -\log P_{\Theta+\Delta \Phi + \Delta\mathbf{Z}}(\boldsymbol{I}_t \mid \boldsymbol{I}_{<t})
\end{equation}
where $\Delta\Phi$ and $\Delta\mathbf{Z}$ are the parameters of LoRA adapters and routing respectively. During offline training, the parameters learned by Poly-PRAG are the adapter parameters $\{A_i,B_i\}_{i=1}^{|\mathcal{M}|}$ and the routing matrix $\mathbf{Z}$ in each transformer layer. Table \ref{tab:training_parameters} presents the trainable parameters during offline encoding across different PRAG approaches. 

A practical application situation is handling new documents not in the initial document collection. In Poly-PRAG, since the latent LoRA adapters serve as general adapters, new documents can be integrated by learning how to combine the existing adapters. To achieve this goal, we propose a variant model Poly-z-PRAG, which keeps the pre-trained LoRA adapters fixed and updates only the routing parameters $Z_{\text{new}} \in \mathbb{R}^{|\mathcal{M}|}$. Given a new document $\boldsymbol{d}_{\text{new}}$, we create the augmented QA pairs $\mathbf{A}_{\text{new}} = \left\{\left(\boldsymbol{d}_{\text{new}}^k, \hat{q}_{\text{new}}^j, \hat{a}_{\text{new}}^j\right)\right\}$ and training sample $\boldsymbol{I}_{\text{new}}=[\boldsymbol{d}_{\text{new}}^k; \hat{q}_{\text{new}}^j;\hat{a}_{\text{new}}^j]\in \mathbf{A}_{\text{new}}$. 
The loss function for Poly-z-PRAG is defined as follows:
\begin{equation}
\min_{Z_{\Delta\text{new}}} \sum_{(\boldsymbol{d}_i^k, \hat{q}_i^j, \hat{a}_i^j) \in \mathbf{A}_{\text{new} }} \sum_{t=1}^T -\log P_{\Theta+\Delta \Phi + \Delta Z_{\text{new}}}(\boldsymbol{I}_{\text{new}_t} \mid \boldsymbol{I}_{\text{new}{<t}})
\end{equation}

Table \ref{tab:training_parameters} shows the additional parameters to train in Poly-z-PRAG.
Once the encoding of the new document is complete, we integrate its routing distribution $\Delta Z_{\text{new}}$ with the existing routing matrix $\mathbf{Z}$. This yields an updated routing matrix $\mathbf{Z} \in \mathbb{R}^{(|\mathcal{T}|+1) \times |\mathcal{M}|}$.

\begin{table}[h]
    \centering
    \caption{\textbf{Number of parameters used for PRAG approaches.} $d$ is the input and output dimension of the training parameters (assumed identical). $r$ is the rank in the LoRA, where $r \ll d$. $L$ is the number of transformer layers. Note that $\mathcal{O}(|\mathcal{T}| \times |\mathcal{M}|)$ is considered negligible compared to other parts. $\mathcal{F}_{\phi}$ is the hypernetwork used in DyPRAG.}
    \label{tab:training_parameters}
    \renewcommand{\arraystretch}{1.3} 
    \begin{tabular}{lcl}
        \toprule
        \multirow{2}{*}{\textbf{Model}} & \multicolumn{2}{c}{\textbf{Offline Training}} \\
        \cmidrule(lr){2-3}
                       & \textbf{Trainable Par.} & \textbf{Complexity} \\
        \midrule
        PRAG           & $(2dr \times |\mathcal{T}|)\times L $ & $\mathcal{O}(dL|\mathcal{T}|)$ \\
        DyPRAG         & $(2dr \times |\mathcal{T}|  + \mathcal{F}_{\phi} )\times L$ & $\mathcal{O}(dL|\mathcal{T}|+\mathcal{F}_{\phi}$) \\
        Poly-PRAG      & $(2dr \times |\mathcal{M}| + |\mathcal{T}| \times |\mathcal{M}|) \times L$ & $\mathcal{O}(dL|\mathcal{M}|)$ \\
        Poly-z-PRAG & $|\mathcal{M}|\times L$ & $\mathcal{O}$($|\mathcal{M}|$)\\
        \bottomrule
    \end{tabular}
\vspace{-2mm}
\end{table}

\subsection{Online Inference}
\label{sec:online_inference}


During online inference, Poly-PRAG adopts a \textit{retrieve–activate–generate} paradigm for online inference. Let $\mathcal{D}=\{\boldsymbol{d}_1,\boldsymbol{d}_2,...,\boldsymbol{d}_{|\mathcal{T}|}\}$ denote the set of external candidate documents. As illustrated on the right side of Fig.~\ref{fig:poly_prag}, the process consists of three steps.
(1) Given an input query $q$, a retriever $\mathcal{R}$ selects the top-$k$ most relevant documents: $\mathcal{D}_k(q) = \{ d_{i_1}, d_{i_2}, \dots, d_{i_k} \}, i_j \in \{ 1, \dots, \mathcal{|T|}\}.$ (2) For each retrieved document $d_{i_j}$ with document ID $i_j$, its routing distribution is obtained from the routing matrix $\mathbf{Z}$. This distribution determines the activation degree of LoRA adapters in a document-conditioned manner, and the selected adapters are merged with the routing distribution accordingly into the LLM. (3) The LLM, augmented with the activated LoRA adapters, then generates the final response.

Compared to traditional PRAG, Poly-PRAG allows documents to remain loaded in memory with the LLM. When a new query $q^t$ arrives, the document-aware routing function directly selects the appropriate latent LoRA adapters for each retrieved document $d_i^t$.

\begin{table*}[h]
\centering
\caption{The experimental results (F1 scores \%) of Poly-PRAG on four knowledge-intensive tasks. The best performance is bolded. The Avg is the average score for each subtask. $^\dagger$ denotes significant improvement with t-test at $p<0.05$ over PRAG.} 
\label{tab:main}
\begin{adjustbox}{width=1.0\textwidth}
\begin{tabular}{l|l|cccc|cc|cc|c}
\toprule
\multirow{2}{*}{\textbf{Base LLM}} & \multirow{2}{*}{\textbf{Method}} & \multicolumn{4}{c}{\textbf{2WQA}} & \multicolumn{2}{c}{\textbf{HQA}} & \multirow{2}{*}{\textbf{PQA}} & \multirow{2}{*}{\textbf{CWQ}} & \multirow{2}{*}{\textbf{Avg.}} \\
\cmidrule(lr){3-6} \cmidrule(lr){7-8}
 & & \textbf{Compare} & \textbf{Bridge} & \textbf{Inference} & \textbf{Compose} & \textbf{Bridge} & \textbf{Compare} &  & \\
\midrule
\multirow{5}{*}{LLaMa3-1B}
& Vanilla & 42.89 & 24.17 & 16.91 & 7.87  & 13.25 & 40.26  & 2.26 & 34.94 & 22.82 \\
& Standard RAG & 41.23 & 26.78 & 22.51 & 10.21  & \textbf{21.38} & 42.46  & 17.65 & 37.39 & 27.45  \\
& P-RAG & 50.20 & 24.34 & 19.11 & 8.24 & 13.65 & 40.90 & 23.58 & 35.86 & 26.99 \\
& DyPRAG & 51.25& 48.15 & 17.35 & 7.54  & 14.05 & 43.90 & 11.33 & 36.86 & \underline{28.80} \\

& Poly-PRAG & \textbf{53.24}$^\dagger$ & \textbf{49.23}$^\dagger$ & \textbf{23.11}$^\dagger$ & \textbf{12.45}$^\dagger$ & 16.78$^\dagger$ & \textbf{44.30}$^\dagger$ & \textbf{24.70}$^\dagger$ & \textbf{37.60}$^\dagger$ & \textbf{32.68}$^\dagger$\\
\midrule
\multirow{5}{*}{Qwen2.5-1.5B}
& Vanilla & 45.74 & 39.06 & 17.04 & 7.27 & 12.18 & 39.46 & 2.87 & 26.47 & 23.76\\
& Standard RAG & 38.75 & 38.84 & 11.87 & 5.68  & \textbf{16.19} & 37.13 & 9.97 & 27.33 & 23.33 \\
& P-RAG & 44.96 & 43.96 & 19.29 & 11.14 & 13.27 & 40.42 & 21.55 & 30.82 &  \underline{28.18}\\
& DyPRAG & 43.03 & 47.20 & 17.04 & 8.55 & 13.72 & 41.39 & 6.64 & 31.94 & 26.19\\
& Poly-PRAG & \textbf{46.74}$^\dagger$ & \textbf{48.74}$^\dagger$ & \textbf{20.04}$^\dagger$ & \textbf{12.03}$^\dagger$ & 15.18 & \textbf{42.46}$^\dagger$ & \textbf{24.03}$^\dagger$ & \textbf{32.85}$^\dagger$ & \textbf{30.26}$^\dagger$ \\
\midrule
\multirow{5}{*}{LLaMa3-8B}
& Vanilla & 54.90 & 55.20 & 24.59 & 14.43  & 19.00 & 45.63  & 7.96 & 42.44 & 33.02 \\
& Standard RAG & \textbf{58.43} & 47.77 & 19.20 & 11.07 & 19.68 & 42.10 & 16.13 & 35.45 & 31.23 \\
& P-RAG & 57.78 & 58.93 & 27.61 & 19.17 & 33.68 & 65.88 & 26.13 & 43.54 & \underline{41.59} \\

& DyPRAG & 57.39 & 56.43 & 25.33 & 18.88 & 24.85 & 58.89 & 13.60 & 41.87 & 37.16\\

& Poly-PRAG & 58.39 & \textbf{59.70}$^\dagger$ & \textbf{28.64}$^\dagger$ & \textbf{20.03}$^\dagger$ & \textbf{34.03} & \textbf{67.04}$^\dagger$ & \textbf{28.20}$^\dagger$ & \textbf{45.37}$^\dagger$ & \textbf{42.68}$^\dagger$ \\
\hline
\bottomrule
\end{tabular}
\end{adjustbox}
\vspace{-2mm}
\end{table*}
\subsection{Computation and Storage Cost Analysis}

In this subsection, we compare the different costs of PRAG approaches and show the advantages of Poly-PRAG. 

\paragraph{Offline Encoding}

A strong limitation of the traditional PRAG is its high offline storage and encoding time cost \cite{prag}. 
DyPRAG \cite{tan2025dynamic} addresses this issue by using a hypernetwork to generate LoRA adapters. After training, only the hypernetwork needs to be stored in memory. However, DyPRAG still requires training a separate LoRA adapter for each document during offline encoding. In addition, training the hypernetwork itself incurs substantial encoding time. 
In contrast, Poly-PRAG encodes the entire document collection using a small set of latent LoRA adapters. The trainable parameters of these models are summarized in  Table \ref{tab:training_parameters}. There is a clear advantage for Poly-PRAG with fewer parameters to train and to store.



\paragraph{Online Computation for Generation}

Let $|d|$ denote the average number of tokens per document, $|c|$ the number of retrieved documents, and $|q|$ the length of the query. In traditional RAG, the inference context length is $|c|*|d| + |q|$. For P-RAG \cite{prag} and DyPRAG \cite{tan2025dynamic}, the inference context contains only the query, i.e., $|q|$, for a single inference. However, when a new query $q_{\text{new}}$ arrives, these methods need unload the previously injected LoRA adapters and load new ones for the retrieved documents. In contrast, Poly-PRAG loads the latent LoRA adapters only once. Adapter selection is handled by the routing function. For a new query $q_{\text{new}}$, the retriever selects relevant documents, and the routing function activates the corresponding adapters that are kept in memory. This design avoids repeated adapter loading and unloading, leading to more efficient inference.

\section{Experiments}

In this section, we describe the experimental setup. We first introduce the datasets and evaluation metrics used in our experiments (Sec. \ref{sec:dataset_eval}). We then present the baseline models (Sec. \ref{sec:baselines}) and the implementation details (Sec. \ref{sec:implementation}).

\subsection{Datasets and Evaluation Metric}
\label{sec:dataset_eval}
To enable a comprehensive comparison with prior PRAG approaches \cite{prag, tan2025dynamic}, we evaluate our method on the same benchmark datasets that cover question-answering requiring multi-hop reasoning and commonsense inference:  2WikiMultiHopQA (2WQA) \citep{ho2020constructing} and HotpotQA \cite{yang2018hotpotqa},  are designed to evaluate the model's multi-hop reasoning ability; PopQA \cite{mallen2022not} assesses factual question answering, challenging the model's capacity to recall accurate knowledge and resolve entity ambiguity; ComplexWebQuestions \cite{talmor2018web} evaluates the model's ability to retrieve and reason over large-scale web content. 2WQA and HotpotQA contain different subsets of reasoning types. Following the settings used in previous PRAG studies \cite{prag, tan2025dynamic}, we report F1 scores (\%) as the evaluation metric. 

\subsection{Baselines}
\label{sec:baselines}
Since our method belongs to the PRAG family, we compare it with the following PRAG approaches:
\textbf{Vanilla} QA using an LLM to generate the answer without any external knowledge. \textbf{Standard RAG} that appends top-retrieved documents to the LLM's input prompt, explicitly instructing the model to reference them when generating answers. \textbf{P-RAG} \cite{prag} injects relevant documents into the LLM's parameters via offline LoRA adapters, reducing reliance on input context length. \textbf{DyPRAG} \cite{tan2025dynamic} shares the same offline one-to-one encoding process as PRAG and adds a hypernetwork to translate the document to LoRA adapters in the online inference. 
For any retrieved document, DyPRAG can transform it into a parametric LoRA without the high storage cost. 
\subsection{Implementation Details}
\label{sec:implementation}


For the external knowledge corpus, we used the Wikipedia dumps. We utilize the BM25 as the retrieval model as used in P-RAG. For each test query, there are three passages that will be retrieved \citep{prag,tan2025dynamic}. In the document augmentation process, each passage is rewritten once, and three QA pairs are generated based on the passage. As a result, each document will correspond to 15 question-answer pairs. 


We deploy Low-Rank Adaptation (LoRA) targeting all primary linear projections across the network. The LoRA layers are configured with an intrinsic rank $r = 4$ and a dropout probability of $0.05$. Optimization is performed using the AdamW optimizer with a base learning rate of $1\times 10^{-4}$ and zero weight decay ($\lambda = 0.0$). To simulate a global batch size of $16$ under localized hardware memory constraints, training is executed using a micro-batch size of $1$ paired with gradient accumulation across $16$ successive forward steps.

A core element of our approach is the dynamic routing of inputs, governed by the \texttt{poly\_router}  mechanism. In this setup, the number of latent LoRA adapters is set to $|\mathcal{M}| = 20$. This guarantees that the backbone Transformer layers remain completely frozen, while optimization pressure is isolated to the newly introduced LoRA parameters ($A$ and $B$ matrices), the modular logit routing banks, and the core routing selection tensors. Text sequences exceeding the $2048$ token ceiling are left-truncated to preserve the architectural focus on end-of-sequence prompt dynamics, while the maximum generation length is restricted to $64$ tokens during pipeline evaluation across all evaluation tasks. For the Gumbel-sigmoid function, we use a temperature equal to 1.




\subsection{Results}

\begin{table*}[t]
\centering
\caption{Ablation study of Poly-PRAG. We remove the routing function in each Poly-PRAG layer and assign uniform weights to all latent LoRA adapters.} 
\label{tab:routing_analysis}
\begin{adjustbox}{width=1.0\textwidth}
\begin{tabular}{l|l|cccc|cc|cc|c}
\toprule
\multirow{2}{*}{\textbf{Base LLM}} & \multirow{2}{*}{\textbf{Method}} & \multicolumn{4}{c}{\textbf{2WQA}} & \multicolumn{2}{c}{\textbf{HQA}} & \multirow{2}{*}{\textbf{PQA}} & \multirow{2}{*}{\textbf{CWQ}} & \multirow{2}{*}{\textbf{Avg.}} \\
\cmidrule(lr){3-6} \cmidrule(lr){7-8}
 & & \textbf{Compare} & \textbf{Bridge} & \textbf{Inference} & \textbf{Compose} & \textbf{Bridge} & \textbf{Compare} &  & \\
\midrule
\multirow{2}{*}{LLaMa3-1B}

& Poly-PRAG & 53.24 & 49.23 & 23.11 & 12.45 & 16.78 & 44.30& 24.70& 37.60 & 32.68 \\
&w/o routing & 41.03 & 25.47 & 17.17 & 7.33 & 11.24 & 41.03 & 11.13 & 32.04 & 23.31 \\
\midrule
\multirow{2}{*}{Qwen2.5-1.5B}
& Poly-PRAG & 46.74 & 48.74 & 20.04 & 12.03 & 15.18 & 42.46 & 24.03 & 32.85 & 30.26 \\
&w/o routing & 44.03 & 38.43 & 17.01 & 8.44 & 11.24 & 40.01 & 15.43 & 26.03 & 25.08 \\
\midrule
\multirow{2}{*}{LLaMa3-8B}
& Poly-PRAG & 58.39 & 59.70 & 28.64& 20.03 & 34.03 & 67.04 & 28.20 & 45.37 & 42.68 \\
&w/o routing & 55.04 & 56.03 & 24.39 & 17.01 & 21.02 & 54.03 & 14.29 & 42.39 & 35.53 \\
\hline
\bottomrule
\end{tabular}
\end{adjustbox}
\vspace{-2ex}
\end{table*}

Table \ref{tab:main} presents an evaluation of different RAG methods -- Vanilla, Standard RAG, P-RAG, DyPRAG, and Poly-PRAG -- across three distinct base Large Language Models (LLMs). We show F1 scores on different datasets: 2WQA (Compare, Bridge, Inference, Compose), HQA (Bridge, Compare), PQA, and CWQ, as well as the average score (Avg). Note that the numbers reported in this table may be different from those in the \cite{prag} due to small differences in data augmentation and LoRA training across running environments.


Across all three backbone models, Poly-PRAG consistently delivers the strongest overall performance in most cases. In particular, for smaller models (LLaMA3.2-1B and Qwen2.5-1.5B), Poly-PRAG shows particularly clear advantages, with substantial improvements in average performance and dominant gains on multi-hop reasoning tasks, especially bridge-style and inference-heavy question answering. These results indicate that Poly-PRAG is highly effective at compensating for limited model capacity by improving reasoning and evidence aggregation. For the larger LLaMA3-8B model, Poly-PRAG remains the best-performing method overall, achieving the highest average score and strong results across diverse QA categories. 

While Poly-PRAG excels broadly, we observe task-specific strengths among different RAG variants: Standard RAG performs best on comparison-style questions in 2WQA, and PRAG shows advantages on certain composition and bridge tasks. This suggests that as model capacity increases, different retrieval–generation strategies may specialize in different reasoning patterns. We can also observe that DyPRAG underperforms P-RAG in several cases, confirming that using a single internal hidden state as input to the translator to generate LoRA parameters may not be sufficient. In contrast, Poly-PRAG does not suffer the same problem. Overall, these results highlight Poly-PRAG’s robustness and scalability across model sizes, with especially pronounced benefits for complex multi-hop reasoning, while also revealing nuanced task-level trade-offs among advanced RAG methods.

\subsection{Ablation Study}

For the ablation study, we analyze several mechanisms involved in Poly-PRAG. 

\subsubsection{Routing analysis}
\label{sec:ablation_routing}

We examine the importance of the routing strategy of Poly-PRAG. Table \ref{tab:routing_analysis} presents an ablation study of Poly-PRAG in which the routing function is replaced by uniform weights during offline encoding, forcing each document to activate all adapters equally. Across all base LLMs and datasets, removing routing consistently degrades performance, demonstrating the critical role of routing in Poly-PRAG. For LLaMa3-1B, disabling routing leads to substantial drops across all tasks, with the average score decreasing from 32.68\% to 23.31\%. The degradation is particularly pronounced on reasoning-intensive subtasks such as 2WQA Inference and Compose, as well as on downstream benchmarks like PQA. A similar trend is observed for Qwen2.5-1.5B, where the average performance drops from 30.26\% to 25.08\% when routing is removed. Notably, large declines appear in multi-hop and comparison-style tasks (e.g., 2WQA Bridge and CWQ), indicating that uniform adapter activation hinders the model’s ability to adapt to different document-specific reasoning requirements. For the larger LLaMa3-8B model, although the absolute performance remains higher than smaller models, the impact of removing routing is still substantial, with the average score decreasing from 42.68\% to 35.53\%. 


Overall, these results confirm that document-aware routing is a key component of Poly-PRAG. By enabling the selective activation of latent LoRA adapters, routing allows the mixture of adapters  to better align with the document, 
leading to consistent and substantial performance gains across model scales and evaluation settings.

\subsubsection{Number of latent LoRA adapters}
\label{sec:latent_lora}

We can set different numbers of LoRA adapters in Poly-PRAG. A critical question is: \textit{How many LoRA adapters are sufficient to effectively represent all the documents?}. In this subsection, we examine the impact of the number of LoRA adapters. 

\begin{figure*}[htpb]
        \centering
        \begin{subfigure}
            {0.24\textwidth}
            \centering\includegraphics[width=\textwidth]{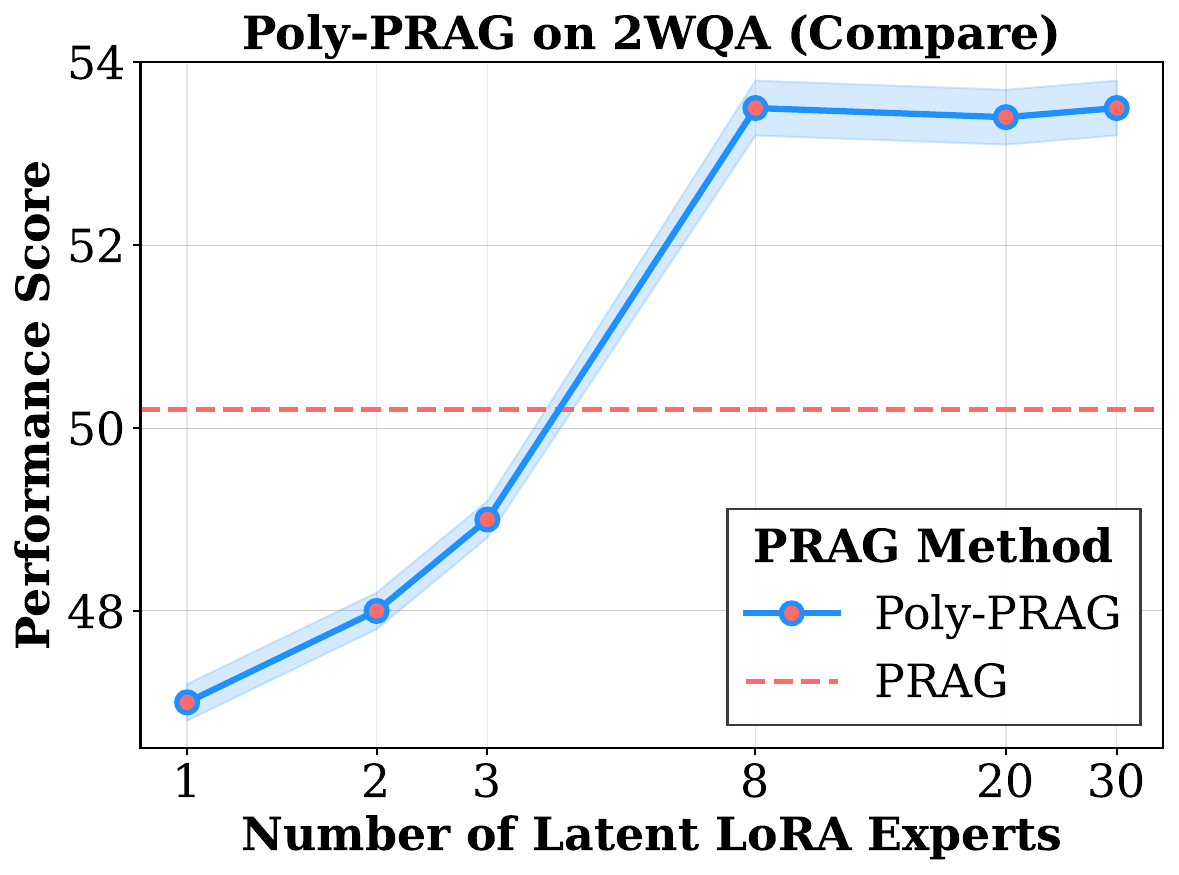}
        \end{subfigure}
        \begin{subfigure}
            {0.24\textwidth}
            \centering
    \includegraphics[width=\textwidth]{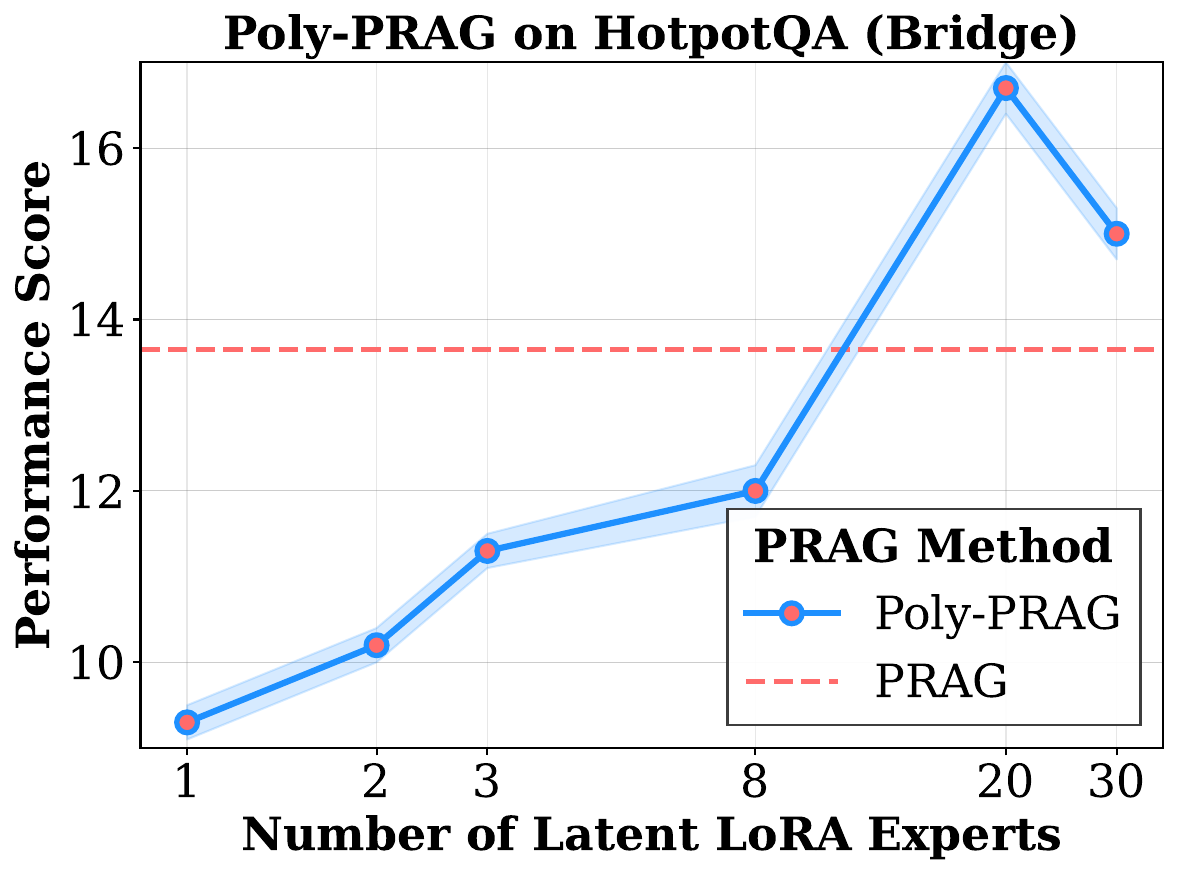}
        \end{subfigure}
        \begin{subfigure}
            {0.24\textwidth}
            \centering
        \includegraphics[width=\textwidth]{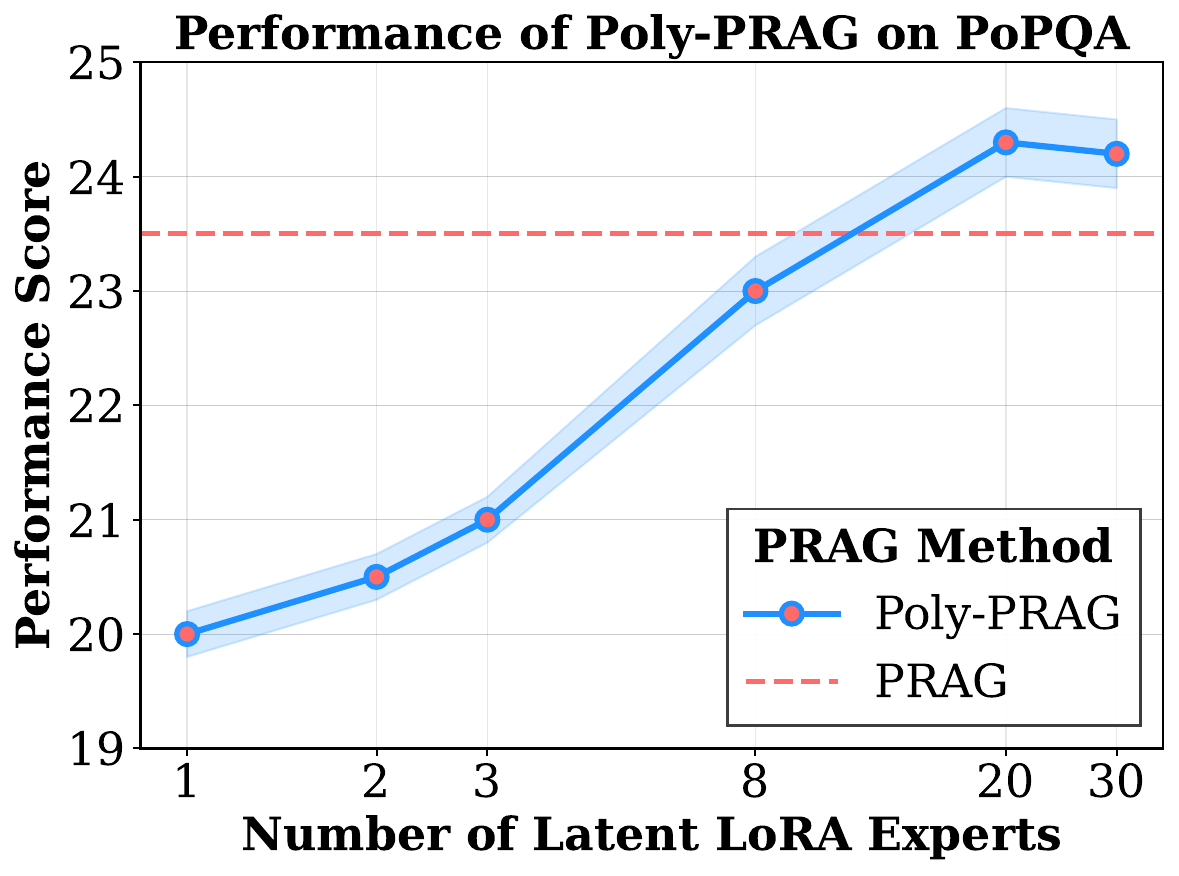}
        \end{subfigure}
        \begin{subfigure}
            {0.24\textwidth}
            \centering
        \includegraphics[width=\textwidth]{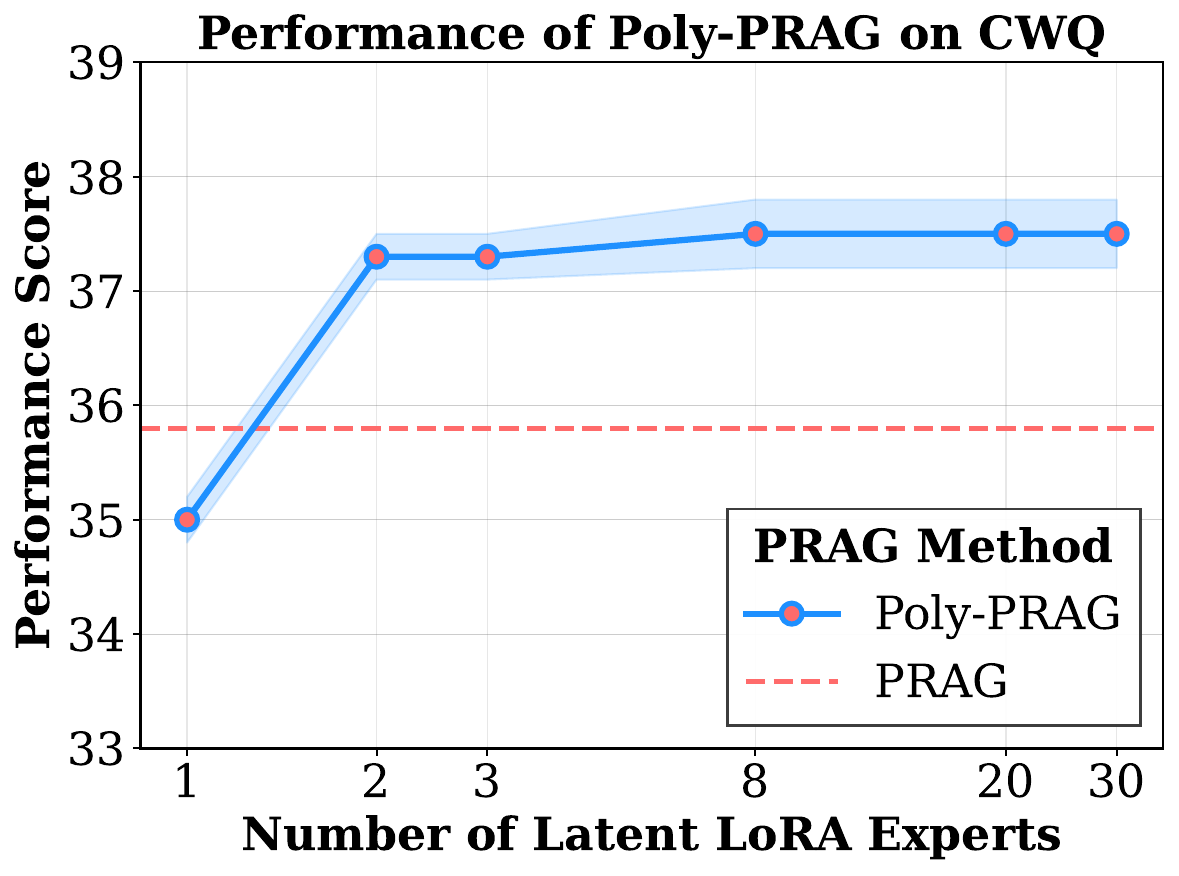}
        \end{subfigure}
        \caption{Analysis of the number of latent adapters of Poly-PRAG based on the LLama3.2 1B.}
        \label{fig:latent_loras}
    \vspace{-2mm}
    \end{figure*}


In Fig. \ref{fig:latent_loras}, we vary the number of LoRA adapters in  \{1, 2, 3, 8, 20, 30\} and show the resulting F1 Score of Poly-PRAG across benchmarks. The optimal number of latent adapters varies across tasks due to differences in task complexity. For 2WQA, the F1 score rises quickly and peaks at 20 adapters, then slightly drops or stays stable at 30. For HQA, the F1 score also peaks at 20 adapters and drops at 30. For PQA, the F1 score increases steadily as more adapters are added, with the best result at 30 adapters. For CWQ, performance saturates early and reaches its peak at 8 adapters, with little change at 20 or 30 adapters. The clearest case of performance drop appears in HQA, where the F1 score decreases by about 1.7 points when increasing from 20 to 30 adapters. This variability highlights that the optimal capacity for knowledge specialization must be carefully tuned for the specific downstream task. However, 20 adapters seem a reasonable choice for all the datasets. Notice, however, that the experiments are performed on only 300 documents per dataset.  More adapters may be required when the whole datasets are used.

\subsubsection{Document representation analysis}

\begin{figure}[h]
\vspace{-5pt}
    \centering
    \includegraphics[width=0.8\linewidth]{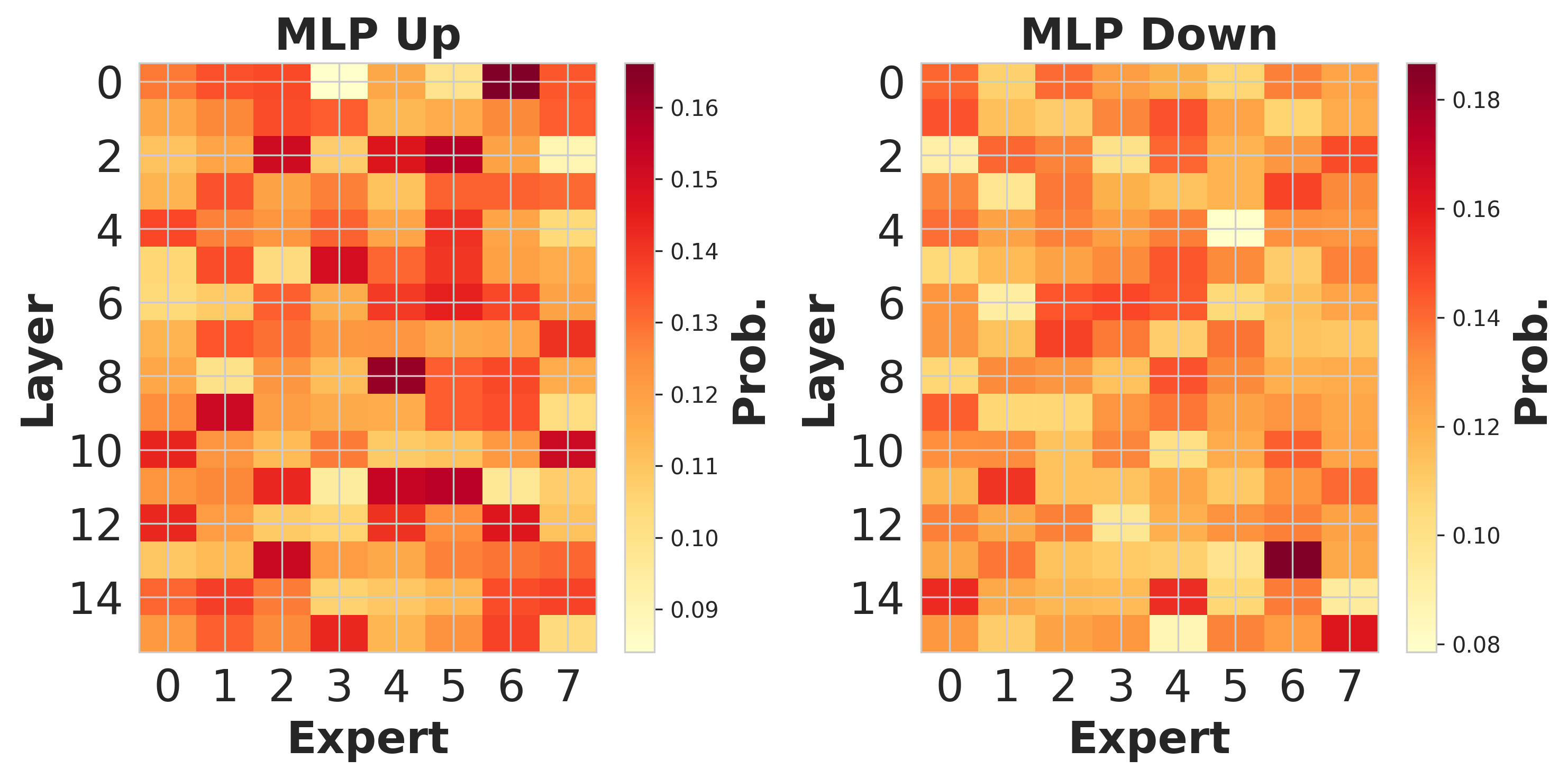}
    \caption{Routing distribution for one document from HQA. We use 8 Latent LoRA adapters in this setting based on LLama3.2-1B. ``Up'' and ``down'' here indicate the different blocks in the feed-forward layer.}
    \label{fig:one_document_visualize}
\end{figure}

In this subsection, we study whether the routing function learns meaningful patterns for document representation.  As shown in the Fig. \ref{fig:one_document_visualize}, we visualize the routing of one document from the HQA dataset. The Poly-PRAG learns structured, layer-dependent routing patterns across LoRA adapters in the up-proj and down-proj blocks of the feed-forward layer. We observe that even within the same transformer layer, LoRA adapters from different blocks learn different routing distributions. The non-uniform and asymmetric adapter utilization indicates meaningful adapter specialization rather than uniform routing, supporting different fits of latent adapters to different documents.

\begin{figure}[h]
\vspace{-10pt}
    \centering
    \includegraphics[width=0.8\linewidth]{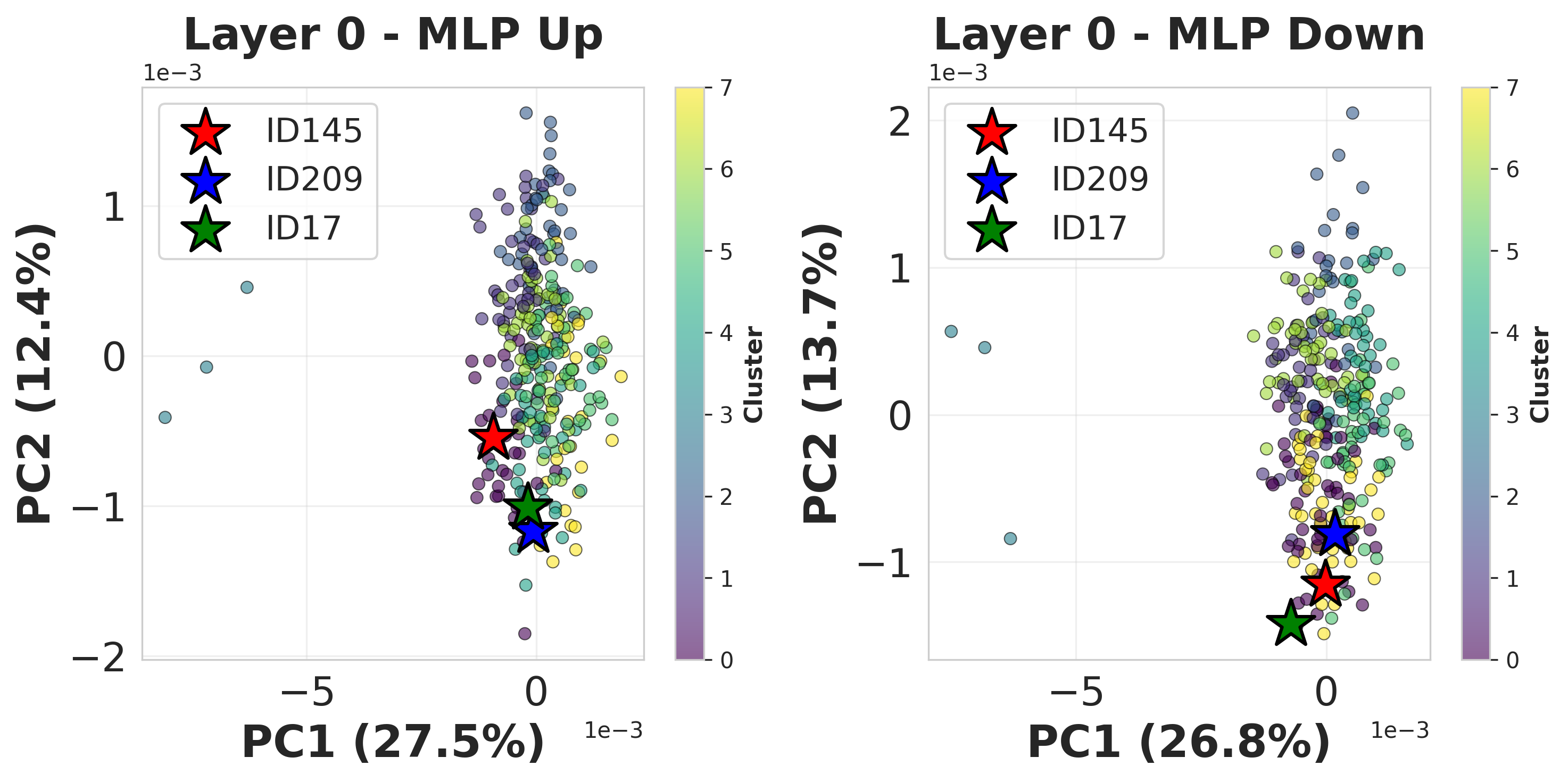}
    \caption{We apply K-means clustering (K=8) to the routing matrix $\mathbf{Z}$ from the MLP layer. We then use PCA to visualize the clustered routing representations.}
    \label{fig:clustering}
\vspace{-2mm}
\end{figure}

We also examine whether the routing function groups similar documents into the same latent LoRA adapters. We extract the routing matrix $\mathbf{Z}$ from MLP layer 0 in Llama3.2-1B for the HQA. In Fig.~\ref{fig:clustering}, we show k-means clusters of the routing matrix. We observe that semantically similar questions are assigned to the same clusters. For example, the question ``What occupations do Charles Burnett and Alan J. Pakula have in common?'' (document ID 145), ``What occupation did Nicholas Christopher and Roger Ebert share?'' (document ID 209) and ``What occupation do Chris Menges and Aram Avakian share?'' (document ID 17) fall into the same cluster. The corresponding adapter(s) would be the one(s) able to answer questions with this pattern.


\subsubsection{Offline Storage Comparison}
\label{sec:storage_comparison}


\begin{table}[h]
\centering
\caption{Comparison of offline storage during encoding for HQA. For P-RAG and DyPRAG, we use the same setting used in the paper \cite{prag,tan2025dynamic}. For Poly-PRAG, we use  8 LoRAs. The ``Temp'' indicates the offline temporary storage during offline encoding. ``Final'' denotes the storage after training.}
\label{tab:storage}
\begin{tabular}{l|l|cc}
\toprule
\multirow{1}{*}{\textbf{Base LLM}} & \textbf{Method} & \textbf{Temp} & \textbf{Final} \\
\midrule
\multirow{3}{*}{LLaMa3.2-1B}
& P-RAG & 12GB & 12GB \\
& DyPRAG & 12GB & \textbf{15MB} \\
& Poly-PRAG & \textbf{42MB} & 42MB\\
\midrule
\multirow{3}{*}{LLaMa3-8B}
& P-RAG & 24GB &  24GB  \\
& DyPRAG & 24GB & \textbf{30MB} \\
& Poly-PRAG & \textbf{218MB} & 218MB\\
\bottomrule
\end{tabular}
\vspace{-2mm}
\end{table}

In Table \ref{tab:storage}, we compare the storage space required for different methods on the HQA dataset. We looked at "Temp" (temporary space needed during the process) and "Final" (space needed after training).
Overall, PRAG is very expensive, requiring gigabytes of storage for both steps. DyPRAG solves the final storage problem, shrinking the model to as little as 15MB, but it still requires a large amount of temporary space (up to 24GB) for training. Notice that the required storage is only for 300 documents in a dataset. For a larger number of documents in a realistic situation, the required storage space would become excessively large to be implementable.

Poly-PRAG offers the best efficiency during the encoding process. It dramatically reduces the temporary storage requirement compared to the other methods. For example, with the LLaMa3.2-1B model, Poly-PRAG needs only 42MB of temporary space, while PRAG and DyPRAG need 12GB. Although DyPRAG has the smallest final size, Poly-PRAG is much lighter to run because it does not require a massive amount of temporary memory.

\subsubsection{Offline Encoding and Online Inference Time}
\label{sec:encoding_time}

\begin{table}[h]
    \centering
    \caption{Offline encoding and online inference time comparison. The backbone model is Qwen2.5-1.5B. We report the average offline encoding time per document on the CWQ datasets. The lowest times among PRAG-based methods are highlighted in \textbf{bold}. For the DyPRAG, we consider the hypernetwork training time during offline encoding and the translation time during online inference.} 
    \label{tab:encoding-time}
    \begin{tabular}{l|c|c}
        \toprule
         \textbf{Method} & \textbf{Offline(s)} & \textbf{Online (s)} \\
        \midrule
        Vanilla & -  & 0.56\\
        Standard RAG & -  & 1.20 \\
        P-RAG  & 15.18  & 0.56\\
        DyPRAG & 19.06  & 0.75\\
        Poly-PRAG & \textbf{4.83}& \textbf{0.40} \\
        \bottomrule
\end{tabular}
\end{table}

Table \ref{tab:encoding-time} compares the speed of different methods using the Qwen2.5-1.5B model. We measured the time needed for ``offline encoding'' and "online inference". Among the PRAG-based methods, Poly-PRAG is the most efficient. It takes only 4.83 seconds for offline encoding, which is much faster than PRAG (15.18 seconds) and DyPRAG (19.06 seconds). During online inference, Poly-PRAG also performs very well. On the CWQ dataset, it is the fastest method overall with a time of 0.40 seconds, beating both the Vanilla model and Standard RAG. 



\subsection{Integrating Additional Documents using Poly-z-PRAG}

To add new documents to Poly-PRAG, we propose a lightweight extension that keeps the base LoRA adapters frozen and fine-tunes only the routing function $\mathbf{Z}$. To evaluate this setting, we reserve $d_{new}$ (3\% of all documents) documents as new documents. We first train Poly-PRAG on documents from the HQA (Compare) dataset. We then encode the $d_{new}$ new documents using Poly-z-PRAG.
The routing function is fine-tuned for 100 steps. As shown in Table \ref{tab:Poly-z}, Poly-z-PRAG achieves performance comparable to Poly-PRAG while adding only a very small number of trainable parameters for the new documents. This result shows the feasibility of reusing a set of trained adapters for new documents. This perspective is particularly interesting for large datasets in practice: A set of fixed adapters can be trained on a sufficiently large number of documents, and other documents can be encoded by reusing the same adapters without retraining them.

\begin{table}[h]
    \caption{Additional Documents integration with Poly-z-PRAG training. We use Qwen2.5-1.5B as the backbone model. ``Params'' indicate the training parameters during encoding.}
    \label{tab:Poly-z}
    \centering
    \begin{tabular}{l|cc}
    \toprule
      Method   &  Params (MB) & F1 score \\
    \midrule
      Poly-PRAG   & 30.29 &  42.46 \\
      Poly-z-PRAG & 0.32 & 41.37 \\
    \bottomrule
    \end{tabular}
\vspace{-3mm}
\end{table}

\section{Conclusion}

We presented Poly-PRAG, a new parametric retrieval augmentation generation framework. Unlike prior approaches that dedicate a LoRA adapter to one document, our method strategically utilizes a pool of shared base LoRA adapters to encode the complete set of documents. This innovative encoding method yields substantial savings in offline storage overhead. Crucially, for online operation, Poly-PRAG incorporates a latent routing function that dynamically selects and activates the relevant adapters on a per-query basis. The proposed model has the ability to capture general knowledge in different adapters, and is agile enough to fit each document through routing. Comprehensive evaluation on various benchmark datasets validates the superior performance of our technique. Ultimately, Poly-PRAG also provides valuable insights into how to effectively and efficiently inject extensive external knowledge into RAG systems.

\section*{Acknowledgment}
This study was supported by the CAISR Health (Center for Applied Intelligent Systems Research in Health, Halmstad University) research profile funded by the Knowledge Foundation (grant 20200208 01H). The computational resources were provided by the National Academic Infrastructure for Supercomputing in Sweden, under project NAISS 2026/3-445 (Scalable and Adaptable Medical LLM (SAM-LLM).

\bibliographystyle{abbrv}

\bibliography{IEEEfull}

\end{document}